\documentclass{article}
\usepackage{arxiv}
\usepackage[utf8]{inputenc} 
\usepackage[T1]{fontenc}    
\usepackage{hyperref}       
\usepackage{url}            
\usepackage{booktabs}       
\usepackage{amsfonts}       
\usepackage{nicefrac}       
\usepackage{microtype}      
\usepackage{lipsum}
\usepackage{graphicx}
\usepackage[T1,OT1]{fontenc}
\usepackage{braket}
\usepackage{soul}
\usepackage{multirow}
\usepackage{graphicx}%
\usepackage{dcolumn}%
\usepackage{bm}%
\usepackage{caption}
\usepackage[labelformat=parens]{subcaption}
\usepackage{float}
\usepackage{romannum}
\usepackage{xcolor}

\title{Patterns for ghost imaging generated by the Ising model}

\author{
 Hamidreza Oliaei-Moghadam \\
  Department of Physics, College of Sciences\\
  Shiraz University\\
  Shiraz 71946-84795, Iran \\
  \texttt{oliaeimoghaddam@gmail.com} \\
   \And
 Chan{\'{e}} Moodley \\
  School of Physics\\
   University of the Witwatersrand\\
   Johannesburg 2000, South Africa \\
  \texttt{chane13.m@gmail.com} \\
  \And
 Mahmood Hosseini-Farzad \\
  Department of Physics, College of Sciences\\
  Shiraz University\\
  Shiraz 71946-84795, Iran \\
  \texttt{ hosseinif@shirazu.ac.ir} \\
}

\begin{document}
\maketitle
\begin{abstract}
\noindent We offer new illumination patterns for imaging in all-digital ghost imaging (GI) systems.
The digital patterns, written as computer generated holograms on spatial light modulators (SLM), are generated by the Ising model, a well-known ferromagnetic model.
In general, Ising patterns are affected by the system temperature and magnetic field. 
For a modest number of measurements in GI, our simulations suggest that low-temperature Ising patterns outperform random patterns. We also experimentally confirm that Ising masks can be used in quantum GI. Moreover, we present a strategy for applying Ising patterns of different temperatures to obtain high-quality images with the number of measurements approximately equal to the number of pixels in the image.
\end{abstract}


\section{Introduction}
GI is an alternative imaging technique which can be used for imaging in both quantum \cite{Wang2012,Bennink2004,Pittman1995} and classical regimes \cite{Valencia2005,Basano2006,Ye2020,Gatti2004}.
GI has received much attention in the last decades due to its advantages over conventional imaging, such as super-resolution \cite{Chen2014,Gong2012,Wang2012,Kong2017}, better image quality in a harsh optical environment \cite{Fan2013,Meyers2011,OsorioQuero2021}, and higher detection sensitivity \cite{Zhang2018,Zhang2019}.
GI utilizes correlation information between two beams, the object and the reference beam, to reconstruct an image of the object or scene. The object beam illuminates the object (scene), then a single-pixel detector monitors all the passing light. Meanwhile, a spatially resolving detector monitors the reference beam that does not see the object. None of the detectors can determine the object's spatial information on their own. Since the two beams are spatially correlated, one can reconstruct the image of the object using information obtained from both the detectors\cite{Shapiro2012,Padgett2017}. The correlation of the object beam and reference beam can be driven by either a quantum source (entangled photons) or a classical source (pseudo-thermal light) \cite{Bennink2004,Gatti2004}. Although GI is thought to have a wide range of applications, many of them have yet to be realized in practice due to the complexity of conventional GI's optical path structure.
\par
Fortunately, Shapiro improved on the classical GI by omitting the reference beam and the spatially resolving detector, resulting in a more straightforward experimental setup. In his novel scheme, the correlation is now between the signal of the single-pixel detector located behind the object and pre-computed patterns that are encoded onto a SLM. This scheme is known as ``computational ghost imaging" (CGI) \cite{Shapiro2008}, and it is closely related to the single-pixel imaging field \cite{Gibson2020,Edgar2018}. The ability to image at wavelengths outside the visible spectrum, where focal plane detector arrays are unavailable or expensive, is a significant advantage of using single-pixel imaging systems \cite{Gibson2017,Duarte2008,Chan2008,Pelliccia2016,Sun2013}.
\par
SLMs were soon employed in quantum GI systems as well \cite{Zerom2011,Howland2013}. In this novel quantum imaging protocol, the object beam (signal photon) interacts with the object, while the reference beam (idler photon) interacts with a SLM. Two-photon detection of the signal and idler photons provide information about the object's similarity to the pattern displayed on the SLM. This method suggests a more cost-effective use of photons for quantum imaging.
\par
Despite all the benefits, GI has a significant drawback. Numerous measurements are required to reconstruct the image, implying that imaging speed is unsatisfactory. 
There have been many attempts to improve GI speed by improving imaging strategies and image reconstruction algorithms \cite{Katz2009,Ferri2010,Chan2009,Zhao2012,Sun2012,Sun2016,Wang2014,Yu2014,Wang2016,Zhang2017,Luo2018,Zhu2019,Wang2021}.
Machine learning algorithms have also been used to improve picture quality using denoising processes and to decrease image reconstruction time \cite{Lyu2017,Rizvi2020,Radwell2019,Moodley2021,rodriguez2020towards}.
\par
In this paper, we introduce new illumination patterns for single-pixel imaging systems. These patterns are generated by the Ising model, a well-known ferromagnetic model \cite{Newell1953,Cipra2018}. We show that employing Ising patterns instead of traditional random patterns can enhance imaging speed. The main statistical characteristics of these patterns are determined by two variables: temperature and magnetization field. Random patterns are, in fact, a subset of Ising patterns in which the temperature tends to infinity.

\section{Theory}
Imaging using the GI technique requires a large number of measurements because each measurement can only retrieve a small portion of the object's spatial information. An expanded light beam with an intensity cross-section of $I_i(x,y)$ illuminates the object in each measurement. Then a bucket detector collects and measures all the passing light through the object. The measured intensity is proportional to 
\begin{equation}
c_i=\int \int dx \;dy \; I_i(x,y)T(x,y),
\label{eqn:total-intensity}
\end{equation}
where $T(x,y)$ is the transmission function of the object.
In traditional GI, a beam with speckle-pattern cross-sections is used and the image estimation (ghost image) of the object can be obtained by \cite{Simon2017,Valencia2005}
\begin{equation}
T'(x,y)=\frac{1}{M}\sum_{i=1}^{M}\left(c_{i}-\left \langle{c} \right \rangle \right) \mathbf{I}_{i}(x,y),
\label{eqn:GIFormula}
\end{equation}
where $\left \langle c \right \rangle=\frac{1}{M}\sum_{i=1}^{M} c_{i}$ is the ensemble average over M measurements.
In SLM-based GI systems, beams with the desired cross-section can be obtained by properly turning the modulator pixels on and off.
For example, speckle patterns could be simulated with random patterns \cite{Katz2009,Sun2012}.
We model each pixel into a magnetic dipole. In this scenario, pixels on or off indicate whether the spin direction is up or down.
As a result, a pixel array can be viewed as a spin array that can be studied using appropriate statistical models. In actual images, spatially nearby pixels tend to have similar values. With this in mind, we employ the Ising model, which takes into account the interaction of the nearest neighbors, forcing nearby spins to orient.
\subsection{Ising model}
The Ising model is a well-known model of ferromagnetism in statistical mechanics.  
The 2D Ising model includes discrete variables $(s_i=\pm1)$ that are defined on a quadratic lattice site and represent the spins.
Assuming the existence of an external magnetic field $H$ and considering the interaction of each spin with its nearest neighbor, the energy of a configuration $\left \{ s_1,s_2,...,s_N \right \}$ is defined as \cite{Pathria2022,Herbut2007}
\begin{equation}
E=-J\sum_{\left \langle i,j \right \rangle}s_is_j-H\sum_{i=1}^{N}s_i,
\label{eg:energyIsing}
\end{equation}
where $J>0$ is spin-spin coupling, and $\left \langle i,j \right \rangle$ denotes that only interactions between neighbors are taken into account.
The partition function is
\begin{equation}
Z=\sum_{\left \{ s_i=\pm1,i=1,...,N \right \}}e^{-\frac{E}{k_BT}},
\label{eg:PartitionIsing}
\end{equation}
where $k_B$ indicates the Boltzmann constant and $T$ denotes the temperature.
From the partition function, one can calculate the magnetization per dipole as follows:
\begin{equation}
m=\left \langle s_j \right \rangle=\frac{1}{Z}\sum_{\left \{ s_i=\pm1,i=1,...,N \right \}}s_je^{-\frac{E}{k_BT}}.
\label{eq:magnetization}
\end{equation}
\par
At very low temperatures ($T \ll J$ ), even if  $H=0$, it is expected that most of the spins will have the same direction .
At very high temperatures ($T \gg J$ ), on the other hand, the interaction between the spins becomes insignificant, and their orientation becomes random $ m  = 0$ . Therefore, temperature can be regarded as a cause of disorder.
In the absence of the magnetic field, the Ising model has global $Z_2$ symmetry under the transformation $s_i \rightarrow  -s_i $ at all sites.
Ising symmetry exists only in the paramagnetic phase with $m = 0$.
When the system phase changes from the paramagnetic phase to the ferromagnetic phase with $m \neq 0$, symmetry breaks spontaneously. The magnetization $m$, which behaves differently in two phases, is called the order parameter, and
the temperature at which the phase transition occurs is called the critical temperature. L. Onsager showed that this phase transition occurs at $k_B T_c /J = 2.269$\cite{Onsager1944}.
\par
\begin{figure}
\centering
     \begin{subfigure}[b]{0.22\textwidth}
         \centering
         \includegraphics[width=\textwidth]{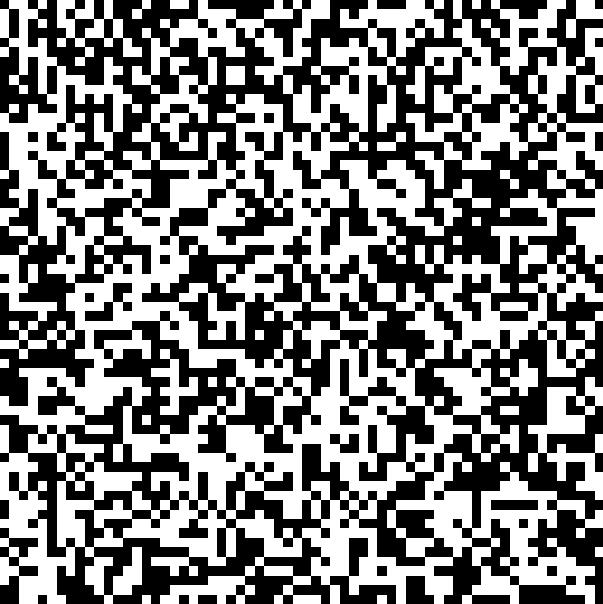}
         \caption{$\beta' = 0$}
         \label{IsingPatterns:1}
     \end{subfigure}
     \hfill
     \begin{subfigure}[b]{0.22\textwidth}
         \centering
         \includegraphics[width=\textwidth]{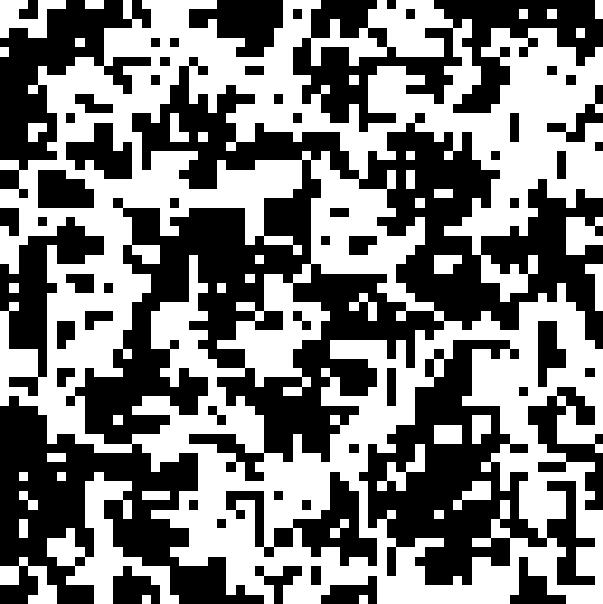}
         \caption{$\beta' = 0.75$}
         \label{IsingPatterns:3}
     \end{subfigure}
     \hfill
    \begin{subfigure}[b]{0.22\textwidth}
         \centering
         \includegraphics[width=\textwidth]{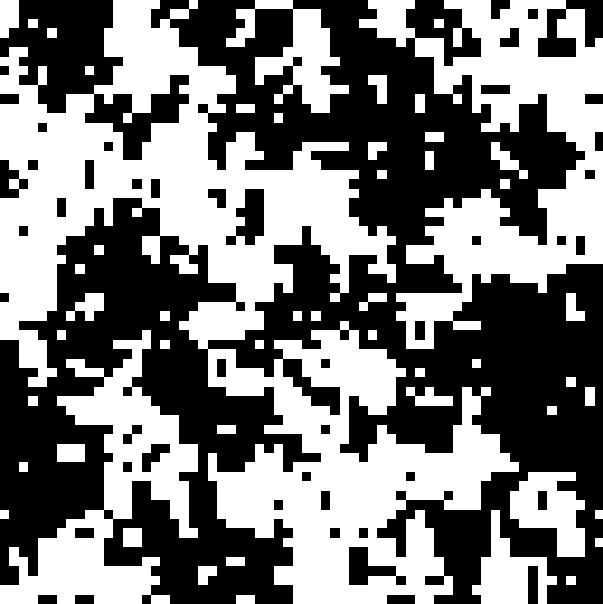}
         \caption{$\beta' = 0.9$}
         \label{IsingPatterns:4}
     \end{subfigure}
     \hfill
    \begin{subfigure}[b]{0.22\textwidth}
         \centering
         \includegraphics[width=\textwidth]{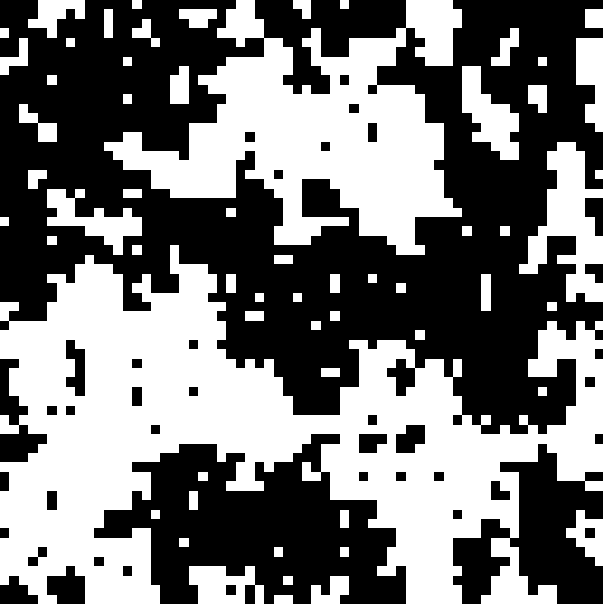}
         \caption{$\beta' \simeq 1$}
         \label{IsingPatterns:5}
     \end{subfigure}
    \caption{\footnotesize{Ising patterns at different value of $\beta'$. The size of the speckles grows as $\beta'$ is increased.}}
\label{fig:IsingPatterns}
\end{figure}

In this paper we work in the regime of $T_c < T < \infty$. For simplicity, it is more convenient to work with the inverse temperature instead of $T$.
By defining the inverse temperature parameter $\beta=\frac{1}{k_BT}$, the desired regime is $0< \beta < \beta_c$. 
The spin configuration is completely random if $\beta=0$ (Fig. \ref{fig:IsingPatterns} (\subref{IsingPatterns:1})), but magnetic domains can be formed if $\beta > 0$.
As $\beta$ increases, so do the number and size of the magnetic domains. When the system reaches its critical point $\beta_c$, the magnetic domains appear at all scales (Fig. \ref{IsingPatterns:5}).
We can scale $\beta$ to $\beta_c$ to get dimensionless parameter $\beta'=\frac{\beta}{\beta_c}$.
To illustrate system behavior by increasing $\beta'$ from 0 to 1, we plotted spin configurations of a square lattice of size $64 \times 64$ for several values of $\beta'=$0, 0.75, 0.9, 1 (Fig. \ref{fig:IsingPatterns}).
These spin configurations are generated using the Metropolis method, the most commonly used Monte Carlo approach for calculating Ising model estimations.
By converting each spin to $s_i \rightarrow \frac{s_i+1}{2}$, we can obtain logical patterns from the array of spins. We refer to these logical patterns as ``Ising patterns."
Ising patterns with $H=0$ and $\beta'<1$ conditions leads to unbiased patterns in pixel values. It should be mentioned that biased Ising patterns can also be
generated if $H \neq 0 $ or $\beta'>1$:
\begin{itemize}
    \item 
    For $\beta'<1$, negative (positive) fields cause sparse (dense) Ising patterns.
    \item
    For $\beta'>1$ and $H = 0 $, the generated Ising pattern is dense or sparse with equal probability.
\end{itemize}
In general, in GI, speckle patterns with small speckle sizes are appropriate for measuring image detail, whereas speckle patterns with large speckle sizes are appropriate for measuring image outline.
Ising patterns have speckles of different sizes and shapes.
This feature of Ising patterns makes it possible to measure the details and outline of the image simultaneously.
\section{Experimental Realization}
\begin{figure}[ht]
\centerline{\includegraphics[width=\linewidth]{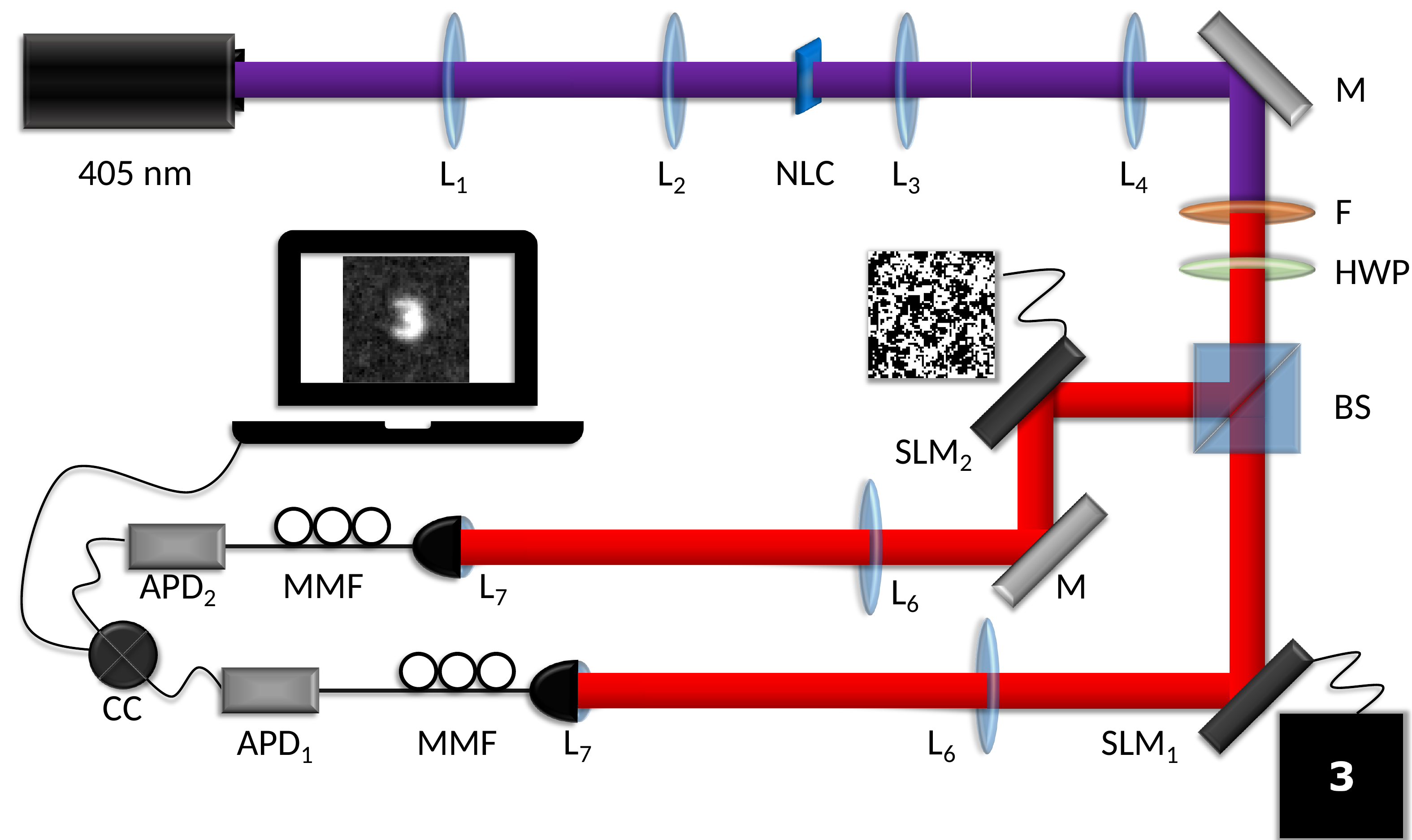}}
\caption{\label{fig:exp}\footnotesize{Schematic diagram of the implemented quantum ghost imaging setup. Entangled bi-photons are produced via spontaneous parametric down-conversion in the non-linear crystal. The entangled photons are spatially separated and imaged on two SLMs. Coincidence measurements are made between both paths, and the coincidences are used to reconstruct an image of the object.}}
\end{figure}

\noindent Fig. \ref{fig:exp} shows a sketch of the quantum ghost imaging optical setup implemented for this work.
Light from a diode laser (wavelength $\lambda = $ 405 nm) was used to pump a temperature-controlled type \Romannum{1} PPKTP non-linear crystal (NLC).
The optimal temperature was set to obtain co-linear emission of entangled bi-photons at 810 nm via spontaneous parametric down-conversion (SPDC). Any unconverted pump photons were filtered out using a band-pass filter centered at 810 nm (F). Vertically polarized down-converted photons emitted from the NLC were then rotated to horizontal polarization by a half-waveplate (HWP). The polarization rotation was done to ensure optimal modulation of the impinging photons by the SLMs situated in each arm. The entangled photon beam was separated into two paths by a 50:50 beam splitter (BS), and each arm was then directed to an SLM, where the required holograms were displayed. The first SLM displayed the digital object to be imaged, while the second SLM displayed the Ising mask patterns. A blazed grating was added to both holograms to separate the first diffracted order from the zeroth, unmodulated order. Photons from each first diffracted order were then coupled into multimode optical fibers (MMFs) connected to avalanche photo-diodes (APDs) for photon detection. Coincidences were measured within a 25 ns window by a photon-counting device (CC). The crystal, SLMs, and fiber entrances are placed in conjugate planes.
After removing the DC component, the image is reconstructed as a linear mixture of the patterns, weighted by the measured coincidences between both arms.
LabVIEW was used to read Ising patterns and show the holograms on each SLM, capture photon correlations, and reconstruct the real-time image using a linear combination of patterns weighted by coincidences.
\section{Results and Discussion}
\begin{figure}[H]
    \centering
    \includegraphics[width=\linewidth]{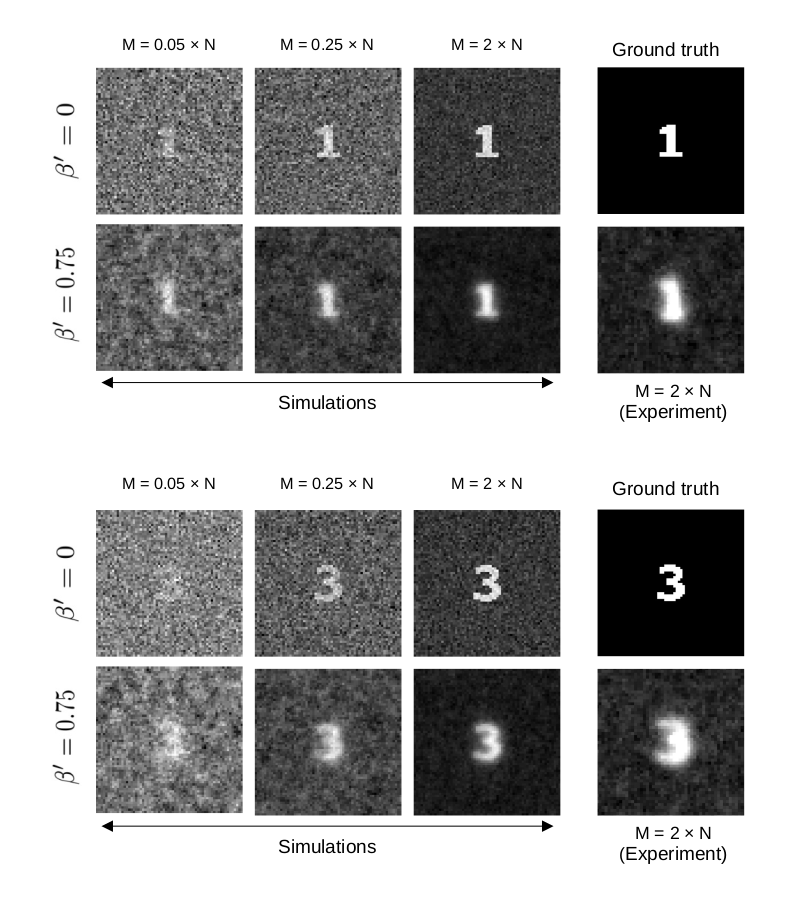}
    \caption{\footnotesize{Experimental and simulation ghost images of two objects. We can see that background of the Ising ghost images with ($\beta'=0.75$) is darker than random ghost images with ($\beta'=0$).}}
    \label{fig:AllResults}
\end{figure}

\begin{figure}[H]
\centering
    \includegraphics[width=\linewidth]{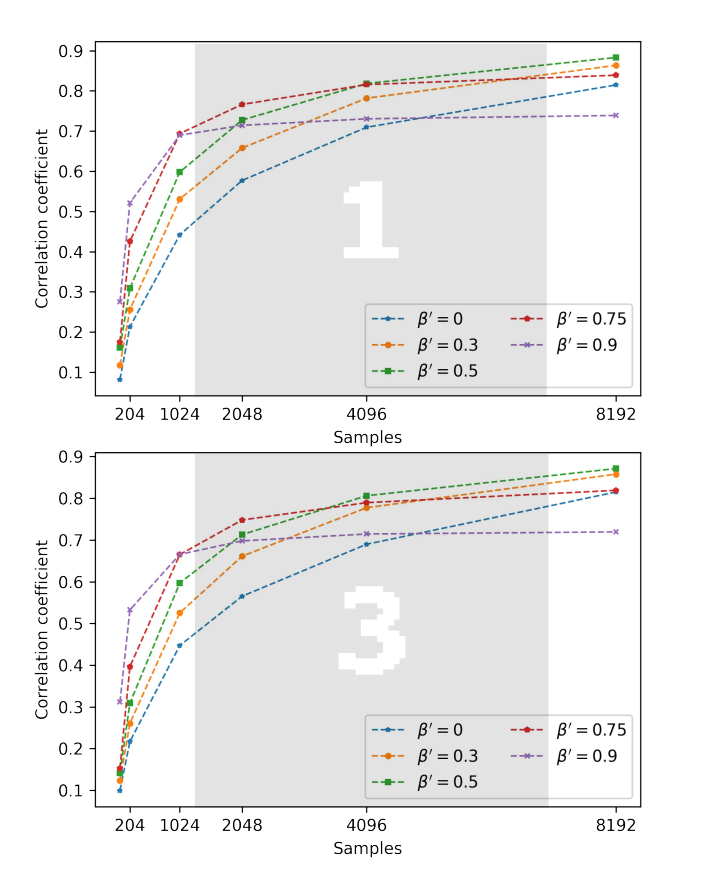}
    \caption{\footnotesize{Quality factor of simulated ghost images. In a small number of samples ( $M < 1024 $ ), pattern set with $\beta'=0.9$ gives a higher correlation coefficient. But when the number of samples increases, pattern sets with lower $\beta$  give better results.}}
\label{fig:corr}
\end{figure}

\begin{figure}[H]
\centering
    \includegraphics[width=\linewidth]{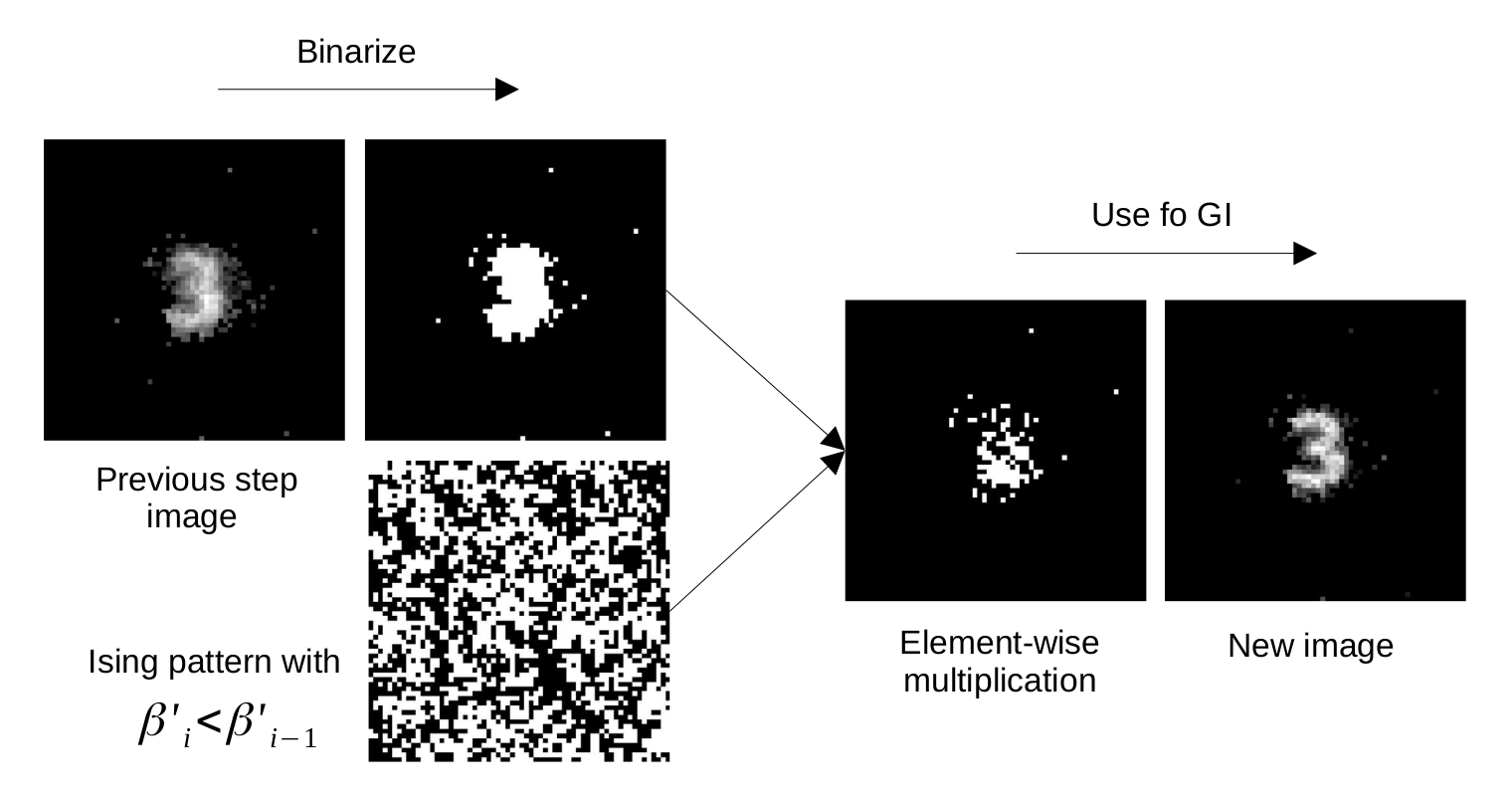}
    \caption{\footnotesize{Diagram for a typical step of applying the multi-$\beta'$ strategy.}}
\label{fig:algo}
\end{figure}

\begin{figure}[H]
\centering
     \begin{subfigure}[b]{0.22\textwidth}
         \centering
         \includegraphics[width=\textwidth]{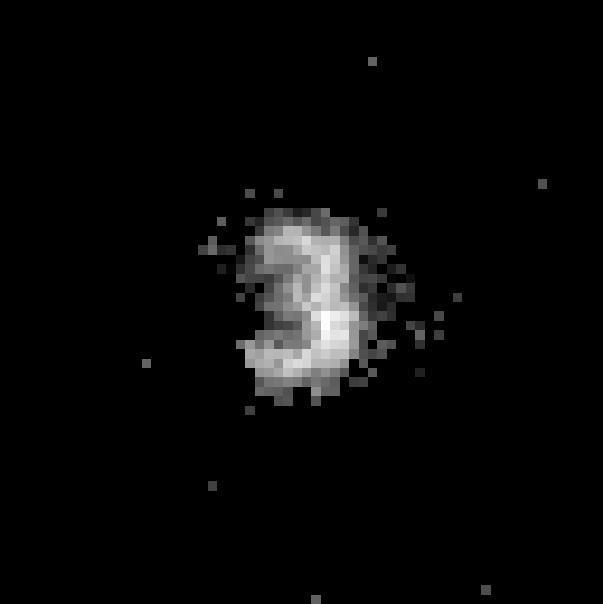}
         \caption{\scriptsize{0.25 $\times$ N}}
         \label{mGIO2:GI0P25}
     \end{subfigure}
     \hfill
     \begin{subfigure}[b]{0.22\textwidth}
         \centering
         \includegraphics[width=\textwidth]{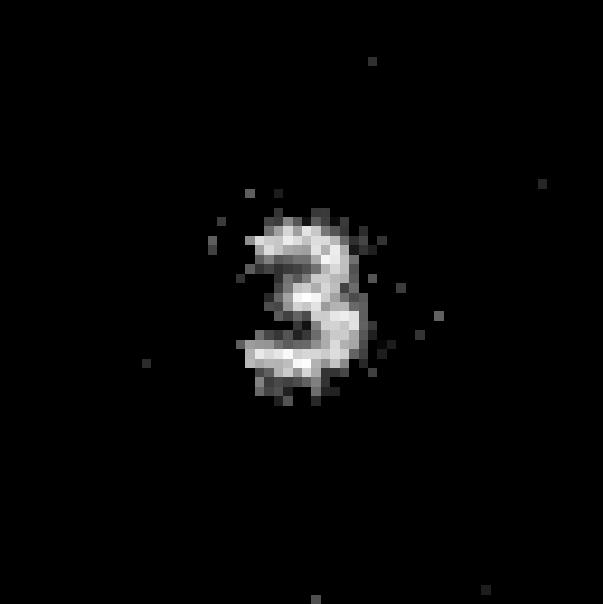}
         \caption{\scriptsize{0.45 $\times$ N}}
         \label{mGIO2:GI0P45}
     \end{subfigure}
     \hfill
    \begin{subfigure}[b]{0.22\textwidth}
         \centering
         \includegraphics[width=\textwidth]{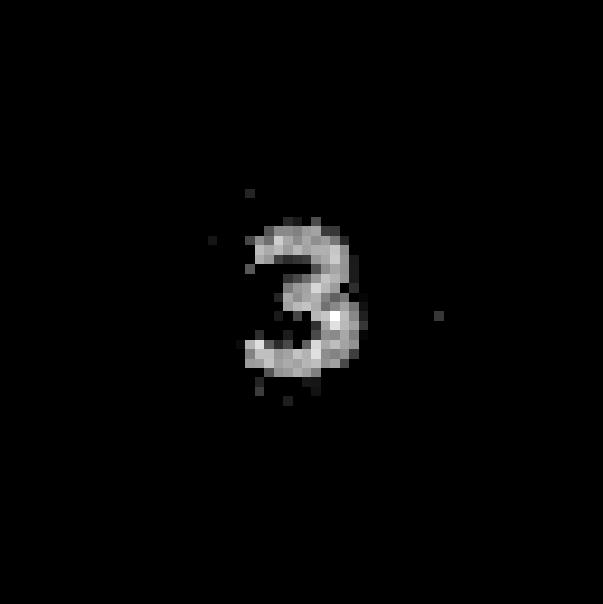}
         \caption{\scriptsize{0.75 $\times$ N}}
         \label{mGIO2:GI0P75}
     \end{subfigure}
     \hfill
    \begin{subfigure}[b]{0.22\textwidth}
         \centering
         \includegraphics[width=\textwidth]{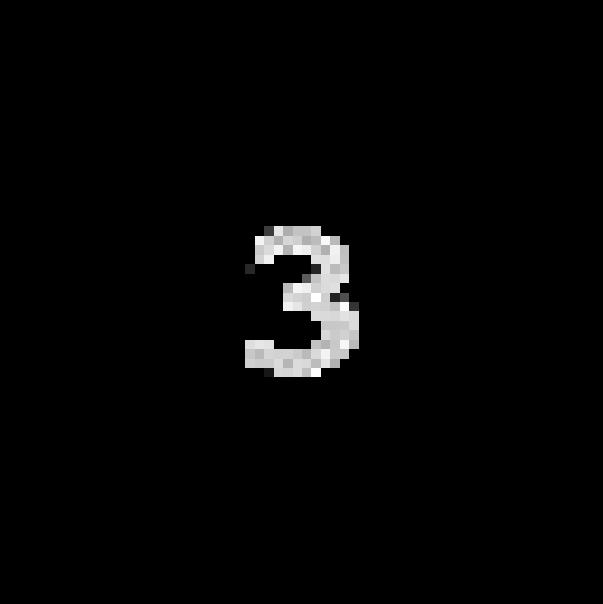}
         \caption{\scriptsize{1.25 $\times$ N}}
         \label{mGIO2:GI0P125}
     \end{subfigure}
    \caption{\footnotesize{Multi-$\beta'$ ghost images. The correlation coefficients for (a)-(d) is equal to 0.797, 0.896, 0.956, 0.993 respectively.}}
\label{fig:GIO2-multi-beta}
\end{figure}

We have generated $64 \times 64$ patterns for several $\beta'=$ 0, 0.3, 0.5, 0.75 and 0.9 as a pre-measurement step.
For simulation results, we used M = 0.01 $\times$ N, 0.05 $\times$ N, 0.25 $\times$ N, 0.5 $\times$ N, N, 2 $\times$ N (where N = 64 $\times$ 64) samples to construct ghost images of two different objects, and we used Pearson's correlation coefficient to check the quality of the reconstructed images.
We also ran an experiment in order to evaluate the use of Ising patterns in a quantum ghost imaging optical setup. The experimental results were obtained using M=2 $\times$ N samples of $\beta'=0.75$.
Simulation and experimental ghost images of $\beta'=0.75$ are shown in Fig. \ref{fig:AllResults}. 
We can see from these results that using high $\beta'$ Ising patterns instead of random patterns ($\beta'=0$) leads to ghost images with a darker background and less noise.
Also, the quality factor graphs ( see Fig. \ref{fig:corr} ) show that, given a small number of measurements, Ising patterns with high $\beta'$ are beneficial to random patterns.
However, as the number of measurements increases, the advantage of the high $\beta'$ patterns over random patterns decreases, and the images become blurry. 
For this reason, it is inconvenient to employ patterns of one specific $\beta'$. A multi-$\beta'$ strategy can be utilized instead. One can use higher $\beta'$ patterns to estimate basic information about the scene, such as the size and location of the items within the scene. Then, by using this basic information and lower $\beta'$ patterns, we can generate new masks.
The algorithm for applying the multi-$\beta'$ strategy is as follows:
\begin{enumerate}
    \item 
    Estimate the image using $M_0$ samples of $\beta'_0$ patterns.
    \item
    Binarize the estimated image using the Otsu threshold\cite{Otsu1979}.
    \item
    Generate new patterns by element-wise multiplication of the binarized image and patterns with $\beta'_i<\beta'_0$.
    \item
    Use $M_i$ new patterns to reconstruct the new estimation of the image.
    \item
    Repeat 2-4 with condition $\beta'_{i+1}<\beta'_{i}$ (See Fig. \ref{fig:algo} )
\end{enumerate}
An example of applying this strategy is shown in Fig.
\ref{fig:GIO2-multi-beta}. The number of samples used for this example are
\begin{itemize}
    \item 
    Fig. \ref{fig:GIO2-multi-beta}(\subref{mGIO2:GI0P25}): M = 410 ($\beta' = 0.9$)+ 614 ($\beta' = 0.75$) = 1024 (0.25 $\times$ N)
    \item 
    Fig. \ref{fig:GIO2-multi-beta}(\subref{mGIO2:GI0P45}): M = 1024 + 819 ($\beta' = 0.5$) = 1843 (0.45 $\times$ N)
    \item 
    Fig. \ref{fig:GIO2-multi-beta}(\subref{mGIO2:GI0P75}): M = 1843 + 1229 ($\beta' = 0.3$) = 3072 (0.75 $\times$ N)
    \item 
    Fig. \ref{fig:GIO2-multi-beta}(\subref{mGIO2:GI0P125}): M = 3072 + 2048 ($\beta' = 0$) = 5120 (1.25 $\times$ N)
\end{itemize}

As can be seen from the Figs. \ref{fig:GIO2-multi-beta}(\subref{mGIO2:GI0P75}) and \ref{fig:GIO2-multi-beta}(\subref{mGIO2:GI0P75}), we can achieve a high quality image using $M \simeq N$ samples.
\par
Here, we also like to mention the caustic pattern imaging approach. Caustic patterns can be formed using slowly varying random phase masks and can be used for imaging over a wide variety of length scales and wavelengths \cite{Toninelli2020}.
Caustic patterns become random speckle patterns in the far-field, similar to Ising patterns that become random at the infinite temperature ($\beta'=0$).
\par
When employing a small number of samples, the caustic and Ising approaches are superior to random patterns in different ways.
Caustic patterns can estimate binary images with enhanced edges, while Ising patterns can be used to estimate binary images with
enhanced contrast.
\section{Conclusion}
Ghost imaging offers several advantages over conventional imaging, such as improved image quality in a harsh optical environment, but it has a slow imaging speed. We introduced new pattern sets for SLM-based ghost imaging systems called Ising patterns to speed up the imaging process. Ising patterns are generally determined by two parameters: the magnetic field $H$ and the inverse temperature $\beta$. In this paper, we set $H=0$ and used Ising patterns of several $\beta$ values to reconstruct images of two objects. Our simulations results showed that Ising patterns with higher $\beta$ are beneficial to random patterns when the number of measurements is low and our experimental results verified the usage of Ising masks in quantum GI. We also introduced a multi-$\beta$ strategy for implementing Ising patterns. Using this strategy, we were able to reconstruct high-quality images with $M \simeq N$ samples.
\section{acknowledgments}
The authors would like to thank Professor A. Forbes and Dr.~J.~T.~Francis for helpful discussions and their generous support during the preparation of this manuscript.

\end{document}